\documentstyle[12pt,epsf]{article}
\renewcommand{\theequation}{\arabic{section}.\arabic{equation}}

\begin{document}
%
%
%
\newcommand{\al}{\mbox{$\alpha$}}
\newcommand{\B}{\mbox{$\beta$}}
\newcommand{\G}{\mbox{$\gamma$}}
\newcommand{\Fpi}{\mbox{$F_\pi$}}
\newcommand{\lb}[1]{\mbox{$\bar{l}_#1$}}
\newcommand{\pipi}{\mbox{$\pi\pi\mbox{ }$}}
\begin{flushright}  BUTP-96/3, hep-ph/9601285
\end{flushright}
\begin{center}
{\Large The Chiral Coupling Constants $\lb{1}$ and $\lb{2}$
from \pipi Phase Shifts}
\vskip 2.5cm
{B.~Ananthanarayan\\
P.~B\"uttiker \\ 
Institut f\"ur theoretische Physik
\\Universit\"at Bern, 
CH--3012 Bern, Switzerland\\
}
\end{center}
\vskip 2.0 cm 
\begin{abstract}
A Roy equation analysis of the available $\pi\pi$ phase shift data
is performed with the $I=0$ S- wave scattering length $a^0_0$
in the range predicted by the one-loop
standard chiral perturbation theory.  A suitable dispersive
framework is developed to extract the chiral coupling constants
\lb{1}, \lb{2} and yields \lb{1}$=-1.70\pm0.15$ and \lb{2}$\approx
5.0$.  We remark on the implications of this determination to
(combinations of) threshold parameter predictions of the three
lowest partial waves. 
\end{abstract}
\newpage
\setcounter{equation}{0}
\section{Introduction}
Chiral perturbation theory  \cite{g+l:pl2,g+l:ann,g+l:np} 
provides the low energy effective
theory of the standard model that describes interactions
involving hadronic degrees of freedom and exploits the
near masslessness of the u, d (and s) quarks and the observation
that the pions,  kaons and the $\eta$ could be viewed as the Goldstone
bosons of the spontaneously broken axial
symmetry of massless QCD.  
It is a non-re\-nor\-mali\-zable theory
and involves additional coupling constants that have to be
introduced at each order in the derivative or momentum expansion.
[From here on we confine our attention to the $SU(2)$ flavor
subgroup.]
At leading order, $O(p^2)$, we have the pion decay
constant, $F_\pi \simeq 93\ {\rm MeV}$ in addition to the
mass of the pion, $m_\pi = 139.6\
{\rm MeV}$, henceforth set equal to 1.
As a result, one has for what is arguably the simplest purely
hadronic process of \pipi scattering a prediction for
a key threshold parameter, the $I=0,$ S- wave scattering length
$a^0_0=7/ (32\pi F_{\pi}^2)\simeq 0.16$ \cite{wein}.
There are  ten more coupling constants
at the next to leading order, $O(p^4)$;
four of them,  $\lb{1},\ \lb{2},\ \lb{3}
\ {\rm and} \ \lb{4}$ enter 
the \pipi scattering amplitude \cite{g+l:pl2}.  
As a result, at this order $a^0_0$ (and $a^2_0$) are predicted
in terms of these as well.  One of the cornerstones of standard
chiral perturbation theory
at $O(p^4)$ is a relatively definite prediction
of $a^0_0$ in the range 0.19-0.21.

The coupling constants \lb{1} and \lb{2} have been fixed in the past
from experimental values available in the literature~\cite{nagels}
for the D- wave scattering lengths:
\begin{eqnarray*}
a^0_2=17\pm 3 \cdot 10^{-4}, \ \ a^2_2=1.3\pm 3 \cdot 10^{-4}. \nonumber
\end{eqnarray*}
The  one-loop expressions for these are~\cite{g+l:ann}:
\begin{eqnarray}\label{dlength}
   & \displaystyle  a^0_2  =  \frac{1}{1440 \pi^3 F_\pi^4}
                                       (\lb{1} + 4\lb{2} - \frac{53}{8}),\quad 
     a^2_2  =  \frac{1}{1440 \pi^3 F_\pi^4}
                                       (\lb{1} + \lb{2} - \frac{103}{40}) &
\end{eqnarray}
and yield
\begin{eqnarray*}
\lb{1}=-2.3 \pm 3.7,\quad \lb{2}= 6.0 \pm 1.3.\nonumber
\end{eqnarray*}
These have also been determined from analysis of $K_{l4}$ 
decays~\cite{Riggenbach}
which yield
\begin{eqnarray*}
\lb{1}=-0.7 \pm 0.9,\quad \lb{2}= 6.3 \pm 0.5\nonumber
\end{eqnarray*}
and more recently by estimating higher order corrections
to these decays~\cite{bijnens}
\begin{eqnarray*}
\lb{1}=-1.7 \pm 1.0,\quad \lb{2}= 6.1 \pm 0.5.\nonumber
\end{eqnarray*}
\lb{1} and \lb{2} are coupling constants consistent with the presence of
resonances.  In particular the $\rho$ resonance may make
a significant contribution [see Appendix C in Ref.~\cite{g+l:ann}] and has
also been discussed more recently~\cite{pp}.  Furthermore, 
generalized vector meson dominance~\cite{eckeretal} leads to numerical
values for these consistent with the numbers above.  Tensor
resonances also have been found to make non-trivial
contributions~\cite{toublan}.

The constants \lb{3} has been estimated from the analysis of
$SU(3)$ mass relations which yields~\cite{g+l:ann}:
\begin{eqnarray*}
\lb{3}=2.9\pm2.4.
\end{eqnarray*}
The variation of $a^0_0$ is essentially equivalent to the variation of \lb{3}.
While here
$\lb{3}$ would have to be $-70$ in order to accommodate
$a^0_0=0.26$, 
there is an extended framework which
re-orders the chiral power counting in order to accommodate
such large values of $a^0_0$ modifying
even the tree-level prediction~\cite{KS}.  
Here we confine ourselves to the more predictive
standard chiral perturbation theory whose stringent 
predictions will come under experimental
scrutiny~\cite{JG}.
Thus we note that in our final analysis we cannot claim
an independent determination of \lb{3} via the Roy
equation analysis of this work since $a^0_0$ is varied
anyway in the range predicted by standard chiral perturbation
theory.

The constant $\lb{4}$ enters the one-loop
expression of the relatively
accurately determined value of the ``universal curve''
combination $2a^0_0-5a^2_0$~\cite{MMS} and is also related to the
independent estimates of the scalar charge radius of the
pion.  An $SU(3)$ analysis that has been performed for the
ratio of the kaon and pion decay constants $F_K/F_\pi$ also
provides a measure of this constant $\lb{4}\approx 4.6$~\cite{g+l:np2}.
In the present analysis we treat $a^0_0$ as the only free
parameter to the fit to the data 
and $a^2_0$ is produced as an output corresponding to the
optimal solution of our data fitting.  In particular, the
values we find remain consistent with the universal-curve
band.  Thus we have a determination of the constant $\lb{4}$.
However, we also perform constrained fits with $a^2_0$  computed
from certain universal curve relations that fix $\lb{4}$ a priori.
The influence on the actual numerical fits is found to be minimal
reflecting the weakness of the $I=2$ channel and influences the
determination of \lb{1} and \lb{2} minimally due to reasons we
discuss in subsequent sections.

On the other hand \pipi scattering has been studied
in great detail in axiomatic field theory~\cite{martin}.
(Fixed-t) dispersion relations with two subtractions, a number dictated by
the Froissart bound, have been rigorously established in the axiomatic 
framework. The properties of crossing and analyticity  have been exploited
in order to establish the Roy equations \cite{roy,BGN}, a system of
integral equations for the partial waves. The Roy equations 
have been the basis of analysis of phase shift data~\cite{BFP}  
and a knowledge of the threshold parameters
involved in \pipi scattering has been obtained.  
Best fits to Roy equation analysis of data are obtained with
$a^0_0=0.26\pm 0.05$ \cite{f+p}. Note that the D- wave
scattering lengths cited earlier have also been extracted from Roy equation
analysis.   The
properties of analyticity, unitarity and crossing and positivity
of absorptive parts have also been shown to produce non-trivial
constraints on the magnitudes of a
certain combination of \lb{1} and \lb{2}\cite{ATW2}.

Here we report on a direct determination of the chiral coupling
constants from the existing phase shift data itself by performing
a Roy equation fit to it when $a^0_0$ is confined to the
range predicted by chiral perturbation theory.  The chiral amplitude 
is rewritten
in terms of a dispersive representation with a certain number
of effective subtractions consistent with $O(p^4)$ accuracy,
where the subtraction constants are expressed in terms of the chiral
coupling constants.  The fixed-t dispersion relations of
axiomatic field theory are also rewritten in a form whereby
a direct comparison can be made with the chiral dispersive
representation, while the effective subtraction constants
are now computed in terms of physical partial waves produced
by the Roy equation fit, the input value of $a^0_0$ and the
resulting value of $a^2_0$ generated by the fit.
In most of our treatment we limit ourselves to a certain approximation
where we account for the absorptive parts
of $l\geq 2$ states only through 
the ``driving terms'' of the Roy equations
for the S- and P- waves.

Furthermore, we also
perform an analysis of certain threshold parameters computed from the Roy 
equation fits which reveals the magnitude of $O(p^6)$ corrections
their one-loop predictions are expected to suffer from.  
The dispersive framework can be extended to meet the needs of
two-loop chiral perturbation theory and used to pin down the associated
coupling constants~\cite{gentwoloop,bijnens+}.  The work reported
here summarizes the first stage of our computations and is presently
being extended to meet the needs of the two-loop computation 
of~\cite{bijnens+}.

\setcounter{equation}{0}
\section{\pipi Scattering to $O(p^4)$ in chiral perturbation theory and
the Roy equation solutions}
%
The notation and formalism that
we adopt in this discussion follows that of Ref.~\cite{BFP}.
Consider  \pipi scattering:  
\begin{eqnarray*}
	\pi^a (p_a) + \pi^b (p_b) \to \pi^c (p_c) + \pi^d (p_d),
\end{eqnarray*}
where all the pions have the same mass.
The Mandelstam variables $s$, $t$ and $u$ are
defined as
\begin{equation}
	s = (p_a + p_b)^2 ,\quad t = (p_a - p_c)^2 ,\quad t = (p_a - p_d)^2,%
	\quad s+t+u=4.
\end{equation}
The scattering amplitude $F(a,b \to c,d)$
(our normalization of the amplitude differs from
that of Ref.~\cite{BFP} by $32\pi$ and is consistent
with that of Ref.~\cite{g+l:ann,g+m}) can then be written as
\begin{eqnarray*}
  F(a,b \to c,d)  =  \delta_{ab} \delta_{cd} A(s,t,u) + \delta_{ac}
                     \delta_{bd} A(t,s,u) + \delta_{ad} \delta_{bc} A(u,t,s).
\end{eqnarray*}
From  $A(s,t,u)$ we construct the three $s$-channel isospin
amplitudes:
\begin{eqnarray}\label{eq:amp:iso:def}
T^{0}(s,t,u) & = & 3 A(s,t,u) + A(t,s,u) + A(u,t,s), \nonumber\\
T^{1}(s,t,u) & = & A(t,s,u) - A(u,t,s), \\
T^{2}(s,t,u) & = & A(t,s,u) + A(u,t,s). \nonumber
\end{eqnarray}
We introduce the partial wave expansion:
\begin{equation}
  T^I(s,t,u) = %
	32 \pi \sum_{l=0}^{\infty} (2 l + 1) P_l({t-u\over s-4} ) f^I_l(s),
\end{equation}
\begin{eqnarray*}
f^0_l(s)=f^2_l(s)=0, l\ {\rm odd},\quad f^1_l(s)=0, l\ {\rm even}.
\end{eqnarray*}
The unitarity condition for the partial wave amplitudes
$f^I_l(s)$ is:
\begin{equation}
 \mbox{Im}f^I_l(s) = \rho (s) |f^I_l(s)|^2 +\frac{1-(\eta^I_l(s))^2}{4
			 \rho (s)},
\end{equation}
where $\rho (s) = \sqrt{(s-4)/s}$ and $\eta^I_l(s)$ is the elasticity at
a given energy $\sqrt{s}$.
We also introduce the threshold expansion:
\begin{equation}
{\rm Re}f^{I}_l(s)=\left({s-4\over 4}\right)^l\left(a^I_l+b^I_l
\left({s-4\over 4}\right)+\ldots\right),
\end{equation}
where the $a^I_l$ are the scattering lengths and the $b^I_l$ are
the effective ranges, namely the leading threshold parameters.

Chiral perturbation theory at next to leading order gives an explicit
representation for the function $A(s,t,u)$ at $O(p^4)$ \cite{g+l:ann}:
\begin{equation}
  A(s,t,u) = A^{(2)}(s,t,u) + A^{(4)}(s,t,u) +O(p^6),
\end{equation}
with
\begin{eqnarray}
A^{(2)}(s,t,u) & = & \frac{s - 1}{F^2_\pi},\nonumber\\
A^{(4)}(s,t,u) & = & \frac{1}{6 F_\pi^4}\left(3 (s^2 - 1)\bar{J}(s) 
		     \right. \nonumber \\
	       &   & + \left.{[t(t-u) - 2 t + 4 u -2]\bar{J}(t) +
	             (t \leftrightarrow u)}\right ) \nonumber\\
	       &   & + \frac{1}{96 \pi^2 F_\pi^4} \{2(\lb{1}-4/3)(s-2)^2 +
	             (\lb{2}-5/6)[s^2 +(t-u)^2]\nonumber\\
               &   & + 12 s (\lb{4} - 1) -3 (\lb{3} + 4 \lb{4} - 5)\}
\nonumber \\
\mbox{ and }
\bar{J}(z)  & = &  -\frac{1}{16\pi^2}\int^1_0 dx \ln [1-x(1-x)z],
\quad {\rm Im}\bar{J}(z) = {\rho(s)\over 16\pi}\Theta(z-4).  \nonumber
\end{eqnarray}
Note also that at $O(p^4)$ the imaginary parts of the partial
waves above threshold computed ($s>4$) from the amplitude above is:
\begin{eqnarray}\label{chiralabs}
{\rm Im} f^0_0(s) & = & {\rho(s)\over 1024 \pi^2 \Fpi^4}(2s-1)^2 \nonumber \\
{\rm Im} f^1_1(s) & = & {\rho(s)\over 9216 \pi^2 \Fpi^4}(s-4)^2\nonumber \\
{\rm Im} f^2_0(s) & = & {\rho(s)\over 1024 \pi^2 \Fpi^4}(s-2)^2 \\
{\rm Im} f^I_l(s) & = & 0,\quad l \geq 2. \nonumber
\end{eqnarray}
[Note that the chiral power counting enforces the property that the absorptive
parts of the D- and higher waves arise only at $O(p^8)$.]
Furthermore these verify the property of perturbative unitarity, viz.,
when the $O(p^2)$ predictions for the threshold parameters 
$a^0_0=7/(32\pi\Fpi^2),\ a^2_0=-1/(16\pi\Fpi^2),\
b^0_0=1/(4\pi\Fpi^2),\ b^2_0=-1/(8\pi\Fpi^2)$ and
$a^1_1=1/(24\pi\Fpi^2)$ are inserted into the pertinent form of 
the perturbative unitarity
relations:
\begin{eqnarray*}
{\rm Im} f^I_0(s) & = & \rho(s)(a^I_0+b^I_0 (s-4)/4)^2,\ I=0,2 \\
{\rm Im} f^1_1(s) & = & \rho(s) (a^1_1 (s-4)/4)^2.
\end{eqnarray*}
In order to carry out the comparison between the chiral
expansion and the physical scattering data, we first recall
that up to $O(p^6)$, it is possible to decompose $A(s,t,u)$ into
a sum of three functions of single variables as follows~\cite{Stern1}:
\newpage
\begin{eqnarray}\label{eq:a_chi}
& \displaystyle A(s,t,u) = 32\pi\left[
			\frac{1}{3} W_0(s) + \frac{3}{2} (s-u) W_1(t) +
                                \frac{3}{2} (s-t) W_1(u) \right.
 & \nonumber \\
& \displaystyle
         \left.                      + \frac{1}{2} \left( W_2(t) + W_2(u) - 
				\frac{2}{3} W_2(s)\right)\right]. &
\end{eqnarray}
One convenient decomposition of the chiral one-loop amplitude is:
\begin{eqnarray}
W_0(s) & = & {3 \over 32 \pi }
\left[ {s-1 \over F_\pi^2} +
{2 \over 3 F_\pi^4} (s-1/2)^2 \bar{J}(s)  \right. \nonumber \\
& &\left. + {1\over 96 \pi^2 F_\pi^4}
(  2 (\lb{1}-4/3) (s-2)^2 + 4/3 (\lb{2}-5/6)(s-2)^2 \right. \nonumber \\
& &\left. + 12 s (\lb{4}-1)-3(\lb{3}+4 \lb{4}-5))\phantom{{s-1 \over F_\pi^2}}
\hspace{-8mm}\right],  
\end{eqnarray}
\begin{eqnarray}
W_1(s) & = & {1\over 576 \pi F_\pi^4}(s-4)\bar{J}(s),  \\
W_2(s) & = & {1\over 16 \pi}\left[
{1\over 4 F_\pi^4}(s-2)^2 \bar{J}(s)+{1\over 48 \pi^2 F_\pi^4}
(\lb{2}-5/6)(s-2)^2\right], 
\end{eqnarray}
where we note that this decomposition is not unique, with
ambiguities in the real part only.
We observe that the imaginary parts of these functions verify
the relation:
\begin{eqnarray}
{\rm Im} W_I(x)={\rm Im} f^I_0(x),\quad I=0,2 \nonumber \\
{\rm Im} W_1(x)={\rm Im} f^1_1(x)/(x-4), \nonumber
\end{eqnarray}
which may be used to demonstrate the following dispersion relations:
\newpage
\begin{eqnarray}
W_0(s) & = & {-1+72\lb{1}+48 \lb{2}-27\lb{3}-108\lb{4}-864\pi^2\Fpi^2
	     \over 9216 \pi^3 \Fpi^4}  \nonumber \\
       &   & + {59-144\lb{1}-96\lb{2}+216\lb{4}+1728\pi^2\Fpi^2\over 18432
	     \pi^3\Fpi^4}s   \\
       &   & + {-797+360\lb{1}+240\lb{2}\over 184320\pi^3\Fpi^4} s^2 +
	     {s^3\over \pi}\int_4^\infty {dx\over x^3 (x-s)} {\rm Im}
	     f^0_0(x), \nonumber \\
W_1(s) & = & {-s \over 13824 \pi^3 \Fpi^4}+{s^2\over \pi}
	      \int_4^\infty {dx \over x^2 (x-4) (x-s)}{\rm Im} f^1_1(x), \\
W_2(s) & = & {6\lb{2}-5 \over 1152 \pi^3 \Fpi^4} + 
	     {23-24\lb{2} \over 4608 \pi^3 \Fpi^4}s + {60 \lb{2} -77 \over
	      46080 \pi^3 \Fpi^4}s^2  \\
       &   & + {s^3\over\pi} \int_4^\infty
	     {dx \over x^3 (x-s)}{\rm Im} f^2_0(x). \nonumber
\end{eqnarray}
We now reconstruct $A(s,t,u)$
from this dispersive representation for the $W$'s to obtain:
\begin{small}
\begin{eqnarray}\label{Achirecon}
A(s,t,u) & = & {s-1\over \Fpi^2}+{-540+480\lb{1}+960\lb{2}- 180\lb{3}-
		720\lb{4} \over 5760 \pi^2 \Fpi^4}  \nonumber \\
	 &   & -{110+480 \lb{1}-720 \lb{4}\over 5760 \pi^2 \Fpi^4}s -
	       {163-120\lb{1}\over 5760 \pi^2 \Fpi^4}s^2  \nonumber \\
	 &   & +{460-480\lb{2}\over 5760 \pi^2 \Fpi^4}(t+u) -{20u(s-t)
		+20t(s-u)\over 5760 \pi^2 \Fpi^4}  \nonumber \\
	 &   & -{154-120\lb{2}\over 5760 \pi^2 \Fpi^4}(t^2+u^2)  \\
	 &   & +32\pi \left( {1\over 3} {s^3\over \pi}\int_4^\infty
	       {dx \over x^3 (x-s)}\left( {\rm Im}f^0_0(x)-
	       {\rm Im} f^2_0(x)\right)  \right.\nonumber\\
	 &   & \left. +{3\over 2} (s-u) {t^2\over \pi}
	       \int_4^\infty {dx \over x^2 (x-t) (x-4)}
	       {\rm Im} f^1_1(x) \right. \nonumber \\
	 &   & \left. +{3\over 2} (s-t) {u^2\over \pi}\int_4^\infty
	       {dx \over x^2 (x-u) (x-4)} {\rm Im} f^1_1(x) \right.\nonumber\\
	 &   & \left. +{1\over 2} \left( {t^3\over \pi}\int_4^\infty
	       {dx \over x^3 (x-t)}{\rm Im} f^2_0(x) + {u^3\over \pi}
	       \int_4^\infty{dx \over x^3 (x-u)}
	       {\rm Im} f^2_0(x) \right ) \right).\nonumber
\end{eqnarray}
\end{small}
This is seen to be the sum of a polynomial of second degree in
$s$, $t$ and $u$ and a dispersive piece.  The 
problem associated with the non-uniqueness of
the real part of the decomposition into the $W$'s is eliminated by setting
$u=4-s-t$ upon which we obtain a second degree polynomial in
$s$ and $t$:
\begin{small}
\begin{eqnarray}\label{eq:chiral:poly}
\lefteqn{P = \left( \frac{29}{120\pi^2\Fpi^4} - 
	     \frac{\lb{2}}{6\pi^2\Fpi^4}\right)
	     \left(t - {t^2\over 4}  - {s t\over 4}\right)}\nonumber\\
  &   & +\left( -\frac{33}{640\pi^2\Fpi^4} + \frac{\lb{1}}{48\pi^2\Fpi^4} + 
	 \frac{\lb{2}}{48\pi^2\Fpi^4} \right) s^2 \nonumber \\
  &   & +\left(\frac{1}{\Fpi^2} + \frac{97}{960\pi^2\Fpi^4}-\frac{\lb{1}}
         {12\pi^2\Fpi^4}-\frac{\lb{2}}{12\pi^2\Fpi^4}
	 +\frac{\lb{4}}{8\pi^2\Fpi^4} \right) s \\
  &   &  +\left(-\frac{1}{\Fpi^2} - \frac{97}{480\pi^2\Fpi^4}
	 +\frac{\lb{1}}{12\pi^2\Fpi^4}+\frac{\lb{2}}{6\pi^2\Fpi^4}
	 -\frac{\lb{3}}{32\pi^2\Fpi^4}-\frac{\lb{4}}{8\pi^2\Fpi^4} \right).
	 \nonumber
\end{eqnarray}
\end{small}

The Roy equation fit allows us to obtain a representation only for
the S- and P- wave absorptive parts, [with some effects of higher
angular momentum states absorbed
into the driving terms (see Appendix A)].  Thus,
a determination of the physical S- and P- wave absorptive parts,
allows us to construct a set of crossing symmetric 
amplitudes~\cite{BGN,mahoux} from which we extract
a representation for $A(s,t,u)$ (see Appendix A for details):
\begin{eqnarray}\label{Adisprecon}
\lefteqn{A(s,t,u) = \frac{32 \pi}{3}(\gamma^0_0 - \gamma^2_0)s^2 +
	        16\pi \gamma^2_0 (t^2 + u^2) + 
		4\pi \al^{2}_0 (u + t)} \nonumber\\
	 &   & +\frac{8\pi}{3}(\al^{0}_0 - \al^{2}_0)s + 16\pi \B^1_1 t(s-u) +
	       16\pi \B^1_1 u (s-t)  \\
	 &   & + 32\pi \left( {1\over 3} {s^3\over \pi}
\int_4^\infty {dx \over x^3 (x-s)}\left( {\rm Im}f^0_0(x)-
{\rm Im} f^2_0(x)\right) \right.\nonumber\\ 
&  & +\phantom{32\pi}\left. {3\over 2} (s-u) {t^2\over \pi}
\int_4^\infty {dx \over x^2 (x-t) (x-4)} {\rm Im} f^1_1(x) \right.
 \nonumber \\
& & +\phantom{32\pi}\left. {3\over 2} (s-t) {u^2\over \pi}
\int_4^\infty {dx \over x^2 (x-u) (x-4)} {\rm Im} f^1_1(x)  \right.\nonumber\\
& & +\phantom{32\pi}\left.{1\over 2} 
\left( {t^3\over \pi}\int_4^\infty {dx \over x^3 (x-t)}
{\rm Im} f^2_0(x) + {u^3\over \pi} \int_4^\infty{dx \over x^3 (x-u)}
{\rm Im} f^2_0(x) \right) \right),\nonumber
\end{eqnarray}
where
\begin{eqnarray}
   \alpha^I_0 & = & a^I_0 -\frac{4}{\pi}\int_4^\infty\frac{dx}{x(x-4)}
		       {\rm Im}f^I_0(x) + 
	   	       \frac{4}{\pi}\int_4^\infty\frac{dx}{x^2}
		       {\rm Im}f^I_0(x)\quad I=0,2 \nonumber \\
   \gamma^I_0    & = & \frac{1}{\pi}\int_4^\infty\frac{dx}{x^3}
		       {\rm Im}f^I_0(x)\quad I=0,2 \label{abg} \\
   \beta^1_1    & = & \frac{3}{\pi}\int_4^\infty\frac{dx}{x^2(x-4)}
		       {\rm Im}f^1_1(x) \nonumber \\
   \alpha^1_0   & = &\beta^0_1=\beta^2_1=0. \nonumber 
\end{eqnarray}
We are now in a position to compare the two representations
for $A(s,t,u)$ namely the chiral representation eq.(\ref{Achirecon})
and the axiomatic representation eq.(\ref{Adisprecon}).
These are formally equivalent,
with the dispersive integrals in the former
described by chiral absorptive parts whereas in the latter by the physical S-
and P-wave absorptive parts.  For the chiral expansion to reproduce low energy
physics accurately we now require the effective subtraction constants to
match.  Once more setting $u=4-s-t$ yields the polynomial piece of
the representation eq.(\ref{Adisprecon}):
\begin{eqnarray}\label{eq:disp:poly}
P & = & -128\pi(\B^1_1 + \G^2_0)( t - {t^2\over 4} - {s t\over 4})+
	\frac{16\pi}{3}(2 \G^0_0 + \G^2_0 - 3 \B^1_1) s^2 
	\nonumber\\
  &   &  + 8\pi ({\al^0_0\over 3} - {5\over 6}\al^2_0 + 8\B^1_1 - 16\G^2_0)s
+ 16\pi (\al^2_0 + 16 \G^2_0).
\end{eqnarray}
A straightforward comparison of eq. (\ref{eq:chiral:poly}) and
eq. (\ref{eq:disp:poly}) 
yields
explicit expressions for \lb{1}, \lb{2}, \lb{3} and \lb{4}.
In particular we have for \lb{1} and \lb{2}:
\begin{eqnarray}
\lb{1} & = & 24\pi^2\Fpi^4 ({41\over 960\pi^2\Fpi^4}-{64\pi
\over 3}(\gamma^2_0-\gamma^0_0+3\beta^1_1)),\label{eq:rel1} \\
\lb{2} & = & 24\pi^2\Fpi^4 ({29\over 480 \pi^2 \Fpi^4}+32\pi(
\beta^1_1+\gamma^2_0)).\label{eq:rel2} 
\end{eqnarray}
The actual numerical values we find for \lb{1} and \lb{2}
are reported in a subsequent section.
These have an interesting dependence on the actual physical
phase shifts:  one observes that in eq.(\ref{eq:rel1}), 
the presence of $\gamma^0_0$.  As a result we can anticipate
\lb{1} to be influenced by the input for $a^0_0$.
In contrast, \lb{2}
has no dependence on $\gamma^0_0$ and
depends almost totally on the P- wave contribution via $\beta^1_1$,
as a result of the weakness of the $I=2$ channel which
renders $\gamma^2_0$ negligible in comparison with $\beta^1_1$
(and $\gamma^0_0$).
Since the P- wave happens to be the best determined experimental
quantity, even the Roy equation fits to it are not strongly
influenced by the input value of $a^0_0$.  Thus we expect
a determination of \lb{2} in this manner to be very stable.

The values we find for \lb{3} and \lb{4} from the procedure above
are in rough agreement with the estimates
provided in the introduction, but suffer from the fact that here the
comparison also involves the $O(p^2)$; they are
determined only after large cancelations occur and thus a determination
of these are not expected to be very reliable.  As noted earlier
the determination in this manner of \lb{3} cannot be considered
independent of the input $a^0_0$ and so we do not report it here.
$\lb{4}$ may be determined from the dispersive formulas here.

\setcounter{equation}{0}
\section{Implications to Some Threshold Parameters}
Specific combinations of threshold parameters
appear on the left hand side
when Roy equations for some of the lowest partial
waves (and for instance their energy derivatives)
are evaluated at threshold.
On the right hand sides one has energy integrals
over partial wave absorptive parts.  These
serve as sum rules for such combinations.
Indeed expressions for these sum rules have been
derived even before the Roy equation program
(see e.g.,~\cite{wan:66}).
While there are several inequivalent methods of
obtaining such sum rules, it has been shown that
as long as one is confined to the absorptive parts
of the S- and P- waves alone, each of these methods
would yield identical results for the right hand sides
(for a recent discussion, see~\cite{ATW}).

We consider the Roy equations eq.~(\ref{roy_eq_drv}) in the following limits:
\begin{eqnarray}
& \displaystyle
\lim_{s\to 4+} {d \over ds}\left(12{\rm Re}f^0_0(s)\right),\quad
\lim_{s\to 4+} {d \over ds}\left(24{\rm Re}f^2_0(s)\right) & \nonumber\\
& \displaystyle \lim_{s\to 4+} {18{\rm Re}f^1_1(s)\over (s-4)/4},\quad
\lim_{s\to 4+} {d \over ds}\left({72{\rm Re}f^1_1(s)\over
(s-4)/4}\right).&
\end{eqnarray}
These may then be rearranged to yield the specific combinations of threshold
parameters satisfying:
\begin{eqnarray}
\lefteqn{3 b^0_0 - (2 a^0_0 - 5 a^2_0)  =  \frac{16}{ \pi} \int_4^{\Lambda}
				    \frac{dx}{(x (x - 4))^{2}}%
          	\label{eq:gwd}}  \nonumber\\[2mm]%
     &  & \left\{4 (x - 1){\rm Im}f_0^0(x) - 3{x-2\over2}%
 	  \sqrt{x(x-4)}(a^0_0)^2\right.  \\
     &  & \left. - 9 (x-4) {\rm Im}f_1^1(x) +
	  5 (x-4) {\rm Im}f_0^2(x)\right \} \nonumber\\
     &  & + 12 \lim_{s\to 4+} {d \over ds} {\rm Re\ } d^0_0(s,\Lambda),\nonumber
\\
\lefteqn{6 b^2_0 + (2 a^0_0 - 5 a^2_0) =  \frac{16}{\pi} \int_4^{\Lambda} 
				    \frac{dx}{(x (x -4))^{2}}}%
          			    \nonumber\\[2mm]%
     &   & \left\{ 2 (x-4) {\rm Im}f_0^0(x) + 9 (x-4) {\rm Im}f_1^1(x)  \right.  \\ \label{eq:gwe}
     &   & + (7 x -4){\rm Im}f_0^2(x)\left. -  3(x -2)\sqrt{x(x-4)}
	   (a^2_0)^2 \right\}\nonumber\\
     &   & + 24 \lim_{s\to 4+} {d \over ds}{\rm Re\ } d^2_0(s,\Lambda),
	   \nonumber \\
\lefteqn{18 a^1_1 - (2 a^0_0 - 5 a^2_0) = \frac{16}{ \pi} \int_4^{\Lambda}
                                      \frac{dx}{(x(x-4))^2}%
				      \label{eq:gwa}}\nonumber\\[2mm]%
	&   & \left\{ -2 (x-4)  {\rm Im}f_0^0(s) + 9 (3 x - 4)
               {\rm Im}f_1^1(x)   \right.	       \nonumber \\
& & + \left.   5 (x-4) {\rm Im}f_0^2(x)\right \}+ 
         18\lim_{s\to 4+} {{\rm Re\ }d^1_1(s,\Lambda)\over (s-4)/4}, \\
\lefteqn{18 b^1_1   =  \frac{16}{\pi} \int^{\Lambda}_4 %
	       \frac{dx}{(x(x-4))^3}}\nonumber \\[2mm]
	  &   &\left\{-2(x-4)^3 {\rm Im}f^0_0(x) +%
	        9 (3 x^3 - 12 x^2 + 48 x - 64){\rm Im}f^1_1(x) 
\right. \nonumber \\ \label{eq:a1}
	  &   & + \left. 5 (x-4)^3{\rm Im}f^2_0(x)\right\} + 
		72 \lim_{s\to 4+} {d \over ds}\left(\frac{{\rm Re\ }
		d^1_1(s,\Lambda)}{(s-4)/4}\right).
\end{eqnarray}
From the following limits for the Roy equations:
\begin{eqnarray}
& \displaystyle \lim_{s\to 4+} {{\rm Re}f^0_2(s)\over((s-4)/4)^2},\quad
 \lim_{s\to 4+} {{\rm Re}f^2_2(s)\over ((s-4)/4)^2} & 
\end{eqnarray}
we find expressions of sum rules for the D-wave scattering lengths.
\begin{eqnarray}
a^0_2  & = & \frac{16}{45\pi} \int^{\Lambda}_4 \frac{dx}{x^3(x-4)}%
	        \nonumber \\
	  &   & \left\{(x-4) {\rm Im}f^0_0(x) + 9 (x+4) {\rm Im}f^1_1(x)%
		   + 5(x-4) {\rm Im}f^2_0(x)\right\}  \\
	  &   & +\lim_{s\to 4+} \frac{{\rm Re\ } d^0_2(s,\Lambda)}
	{((s-4)/4)^2},
		 \nonumber \\
a^2_2  & = & \frac{16}{90\pi} \int^{\Lambda}_4 \frac{dx}{x^3(x-4)}%
	        \nonumber \\
	  &   & \left\{2(x-4) {\rm Im}f^0_0(x) - 9 (x+4) {\rm Im}f^1_1(x)%
		   + (x-4) {\rm Im}f^2_0(x)\right\}  \\
	  &   &  +\lim_{s\to 4+} 
\frac{{\rm Re\ }d^2_2(s,\Lambda)}{((s-4)/4)^2}
	.	\nonumber
\end{eqnarray}
The one-loop expressions for the combinations of interest are
\begin{eqnarray}
\label{eq:sr_d}
3 b^0_0 - (2 a^0_0 - 5 a^2_0) & =  & \frac{1}{16 \pi^3 F_\pi^4}
			(2\lb{1}+ 3\lb{2} - \frac{7}{4}), \\
\label{eq:sr_e}
6 b^2_0 + (2 a^0_0 - 5 a^2_0)  & = & \frac{1}{16 \pi^3 F_\pi^4}
			(\lb{1}+ 3 \lb{2} + \frac{5}{8}), \\
\label{eq:sr_a}
18 a^1_1 - (2 a^0_0 - 5 a^2_0)  & = &  \frac{1}{16 \pi^3 F_\pi^4}%
                                       (\lb{2} - \lb{1}- \frac{55}{24}), \\
\label{eq:sr_AI}
  18 b^1_1   & = & \frac{1}{16 \pi^3 F_\pi^4} (\lb{2} - \lb{1}+ \frac{97}{120}) 
\end{eqnarray}
respectively.  Those for the D-wave scattering lengths are
given in eq.(\ref{dlength}).
Since the Roy equations are a consequence of dispersion relations
with two subtractions, we note that in all the above the leading
$O(p^2)$ contribution cancels exactly.  Furthermore, the constants
$\lb{3}$ and $\lb{4}$ are also absent since they accompany only
constant and linear pieces in $s,\ t$ and $u$ in the $O(p^4)$
scattering amplitude.  It would therefore seem
that any two of the six combinations above is suitable for determination
of $\lb{1}$ and $\lb{2}$ since they enjoy the same status as the
D-wave scattering lengths (see eq. (\ref{dlength})).

A careful consideration
of these reveals some interesting characteristics.  
Note for instance that the one-loop
expressions for $18a^1_1-(2 a^0_0-5 a^2_0)$ and for
$18 b^1_1$ depends only on the combination $\lb{2}-\lb{1}$.
If we now evaluate their numerical values for each of our
Roy equation phase shift representations by inserting them
into the sum rules in order to compute
the coupling constants of interest we would
not have a meaningful result since each of the
combinations above represents parallel straight lines
in the $\lb{1},\ \lb{2}$ plane!  
If we consider the further
combination $18 a^1_1 - (2 a^0_0 - 5 a^2_0) - 18 b^1_1$,
its one-loop expression is $-31/160 \pi^3\Fpi^4$.  If we
saturate the sum rule for this combination with $\Lambda=\infty$
and the chiral absorptive parts eq.(\ref{chiralabs})
we reproduce the one-loop result which is a result of
the perturbative unitarity of the chiral expansion.
As a result when the sum rule for this combination is
evaluated with the physical absorptive parts, we get
an answer that is substantially different from its
one-loop expression.   From this we conclude that the
two-loop corrections to this combination must account for
this discrepancy, although we cannot conclude which one of
the elements in the combination receives the correction.
An analogous exercise may be performed for the
$\pi^0 \pi^0$ combinations and in particular for
$30 a_2-b_0$ where $a_2=(a^0_2+ 2a^2_2)/3$
and $b_0=(b^0_0+2 b^2_0)/3$.  A variety of combinations
that arise from totally symmetric amplitudes that
have similar properties has been examined recently~\cite{ATW}.

A final exercise we perform is to compute the
value for $2a^0_0-5a^2_0$ from the Roy equation fit
and then to insert the value of
$\lb{4}$ we find into the one-loop formula for
the same combination:
\begin{eqnarray*}
2a^0_0-5a^2_0 & = &
{3\over 4\pi\Fpi^2}\left(
1+{1\over 8\pi^2\Fpi^2}(\lb{4}+5/8)\right).
\end{eqnarray*}
Once again we observe that
the agreement, while being fair is not exact reflecting
the somewhat large uncertainties in our determination
referred to earlier as well as due to 
the $O(p^6)$ corrections to its one-loop formula.

\setcounter{equation}{0}
\section{Numerical Results and Discussion}
The numerical solutions of the Roy equations are
obtained by using a parameterization similar to the one described
in great detail in~\cite{BFP}.  In this study we have employed
the \pipi scattering from the CERN-Munich experiment
and documented by Ochs~\cite{Ochs} in the region $19\leq s \leq 60$,
 and the high precision $K_{l4}$ experimental
data~\cite{Rosselet} for the phase shift difference $\delta^0_0-\delta^1_1$.
We have devised some checks on our computation in the following
manner:  we have first of all determined the best fit 
to the parameters of the parameterization
referred to above by requiring it to simultaneously yield
the best fit to the data as well as satisfy the Roy equations
in the domain of their validity.  The same parameterization is
considered to be valid in the domain $60\leq s \leq 110$ as well.
Then we have employed the Ochs data in this region and have
required a best fit to the data in $4\leq s \leq 110$ and
that it satisfies the Roy equation in $4\leq s \leq 60$.
In practice the data above $60$ do not influence the parameters
of the fit significantly:  this in turn renders our numerical
results rather stable since many of the quantities we compute
are dominated by the low-energy behavior.
The details of our work will be documented elsewhere~\cite{Paul}.

Our solutions require as an input parameter only
$a^0_0$ and numerically search for those solutions that minimize
the discrepancy with respect to the data.
In Fig.~1 we present our phase shift fits for
$\delta_0^0-\delta^1_1$ obtained for $a^0_0=0.19,$ $
0.20,\ 0.21$ and also for the central value $a^0_0=0.26$.
The final result of the procedure above is an explicit representation
for the lowest partial waves as functions of the energy in terms
of a few parameters that optimally fit the data 
up to $\Lambda=110$ and verifying the Roy equations in their
domain of validity.  

The numerical values of the $\alpha^I_0, 
\ \gamma_0^I,\ I=0,2$ and $\beta^1_1$ may now be
computed from the explicit phase shift representation provided
by the Roy equation fit, and \lb{1} and \lb{2} are extracted from
eq. (\ref{eq:rel1}) and (\ref{eq:rel2})
(the effects arising
from the neglect above the cutoff $\Lambda$ are disregarded;
an unrealistic absolute saturation of the integrands above
$\Lambda$ can lower \lb{1} by $0.13$ and increase
\lb{2} by the same amount), and \lb{4} from the appropriate
comparison of eq.(\ref{eq:chiral:poly}) and eq.(\ref{eq:disp:poly}).

The Roy equation fits
 are also used to compute the combinations of S- and P-wave
scattering lengths and effective ranges of eq.(\ref{eq:gwd})-(\ref{eq:a1}) 
and the high energy tail and higher wave contributions computed
from the driving terms as expressed in them.
For the right hand sides of the sum rules for the D- wave scattering
lengths,  the driving terms for the
Roy equations for the D- waves is not available in the literature.
The S- and P- wave contributions are explicitly accounted for
up to the cut-off and their high energy tail is disregarded
in our tabulation of the results (an absolute (unrealistic)
saturation
of the integrands above $\Lambda=110$ yields an error of
$+0.7\times 10^{-4}$ on $a^0_2$ and an error of $+0.07\times
10^{-4}$ and $-0.22\times 10^{-4}$ on $a^2_2$).  The
only higher wave contribution arises from the $f_2(1270)$~\cite{pdg}
and we
have estimated its contribution from two inequivalent sets of
sum rules for these available in the literature~\cite{wan:66,ATW}
which give almost identical results of $0.54\times 10^{-4}$ for
$a^0_2$ and $0.38\times 10^{-4}$ for $a^2_2$ and have been
included in the final results.

We tabulate our results
in Table 1 and 2.
In Table 1 for a given input of $a^0_0$ we report the
results of our fit for $a^2_0$, $b^0_0,\ b^2_0,\ a^1_1, b^1_1$,
where the last four are obtained from the sum rules of the
previous section and the values of \lb{1}, \lb{2} and \lb{4}
computed from the dispersive relations.
In Table 2 (a), 2 (b) and 2 (c) we give values for the 9 combinations
of threshold parameters of interest computed from the sum rules
and the Roy equation representation and also their one-loop values
obtained by inserting the values of \lb{1} and \lb{2} from the
dispersive analysis.  

An inspection of this table reveals that the values of
the combinations for $18 a^1_1 -(2 a^0_0 - 5 a^2_0)$ in
both columns agree better than the values in the
two columns for  $18 b^1_1$.   Such an agreement is also
seen to be better for $a_2$ than it is for $b_0$.  We conclude
therefore that the determination of \lb{1} and \lb{2} from
the dispersive framework is better consistent with the
one-loop expressions for the scattering lengths than it is
for the effective ranges.  

As a final check we have compared our results obtained
from the Roy equation representation with those obtained
from a simple
analytic parameterization of the type proposed by
Schenk~\cite{Schenk} for the lowest wave phase shifts:
\begin{small}
\begin{eqnarray*}
\tan \delta^I_0(s) & = & \rho(s)\left\{a^I_0+[b^I_0-\frac{a^I_0}
{(s_I-4)/4}+(a^I_0)^3](s-4)/4\right\}\frac{s_I-4}{s_I-s}, \ I=0,2 \\
\tan\delta^1_1(s) &= & \rho(s) \frac{s-4}{4} \{a^1_1 + [b^1_1-
\frac{a^1_1}{(s_\rho-4)/4}](s-4)/4\}\frac{s_\rho-4}{s_\rho-s}.
\end{eqnarray*}
\end{small}
This parameterization employs the postulated normal threshold
behavior and the properties of threshold expansion of the
real part of the partial waves and elastic unitarity and
incorporates features of the \pipi
interaction such as the position of the $\rho$, $s_\rho=
30$, and that the
$I=0$ S- wave passes through $\pi/2$, $s_0=36$ (and an unphysical
$s_2=-16$).   The remainder of the required inputs are obtained from
the first column in each of the Tables 1, 2 and 3 in addition
to the values of $a_0^0$ and $a^2_0$ that correspond to these
and evaluated the $\alpha^I_0, \gamma^I_0, I=0,2$ and $\beta_1^1$
with $\Lambda\approx 50$ ($K$-$\overline{K}$ threshold)
and solved for $\lb{1}$ and $\lb{2}$.   For instance for
the entries of Table 2 (b) we find $\lb{1}=-1.72$ and $\lb{2}=5.2$.
Thus this simple parameterization confirms our numerical findings
to within a surprising $5\%$ accuracy.  However when this
parameterization is inserted back into the sum rules for the
threshold parameters, the agreement is good only to about $15\%$. 

We now remark on contrasting our determinations of \lb{1} and \lb{2}
with those mentioned in the introduction.  Our numbers for \lb{1}
and \lb{2} are
comfortably accommodated in the range that was first obtained from
D- wave scattering lengths in Ref.~\cite{g+l:ann} and more or less
accommodated in the range obtained from the $K_{l4}$ decays.  Our
value for $\lb{2}$ is significantly lower; the
somewhat larger values of the $I=0$ D- wave scattering length 
used in Ref.~\cite{g+l:ann} than the numbers we find appear to
be the cause.   In the present work we have presented a clear
correlation between the input value of $a^0_0$ and the D- wave
scattering lengths which points towards a $10\%$ smaller value
for $a^2_0$ than the central value of $17\times 10^{-4}$ employed
there.  Furthermore, we are working in a specific 
dispersive framework where
the one-loop expressions for the effective subtraction constants
are altogether likely to suffer some higher order corrections,
which we have not attempted to analyze in this work.   Similarly
the other determinations cited are also susceptible to such
corrections and it has not been possible to explain quantitatively
what the origins of the discrepancy are.

\setcounter{equation}{0}
\section{Conclusions}
The coupling constants $\lb{1}$ and $\lb{2}$ of the one loop chiral
expansion has been accurately determined at $O(p^4)$ precision from
a Roy equation analysis of the existing \pipi scattering data with
$a^0_0 \in (0.19,0.21)$ predicted by standard chiral perturbation theory.
A suitable dispersive framework is used to effect a comparison between
the one loop chiral representation
and the Roy equation phase shift representation of the \pipi
amplitudes to obtain 
\begin{eqnarray*}
\lb{1}=-1.70\pm 0.15 \ {\rm and} \ \lb{2}\approx 5.0.
\end{eqnarray*}
The result is consistent with the bounds
obtained on the combination $\lb{1}+2\lb{2}$ in Ref.~\cite{ATW2}.
Certain ambiguities in determinations of these at $O(p^4)$ from 
(combinations
of) threshold parameters is discussed.
The numerical consistency of one-loop results for
those involving certain scattering
lengths with the new values of \lb{1} and \lb{2} and their values
computed from the Roy equation fits is superior to the consistency
for those that involve effective ranges.

\section*{Acknowledgments}  We thank the Swiss National Science
Foundation for support during the course of this work.  It is a
pleasure to thank H. Leutwyler for discussions and crucial insights.
We thank G. Colangelo, J. Gasser and D. Toublan
for comments on the manuscript
and C. D. Frogatt for useful correspondence.

\section*{Note added}  After this work was completed we received
an e-print~\cite{KMSF} dealing with the subject of phase shift data,
sum rules and chiral coupling constants at $O(p^6)$.  
The coupling constants in the first row of Table 4 therein corresponds to
$\lb{1}=-1.36$ and $\lb{2}=5.2$.  This
may be compared with the coupling constants corresponding to our
results for $a^0_0=0.20$.  We find agreement to within a few percent
with the Roy equation solution while the agreement is
very good  for
$\lb{2}$ computed from Schenk's model.  


\appendix
\section*{Appendix A}
\setcounter{section}{1}
\renewcommand{\theequation}{\Alph{section}.\arabic{equation}}
Independent of the dynamics governing the interactions, it
has been rigorously established that
fixed-t dispersion relations with two subtractions may  be
written down for the amplitudes of definite isospin in terms of
unknown t-dependent subtraction functions:
\begin{eqnarray}
T^I(s,t,u) & = &\sum_{I'=0}^2 C_{st}^{II'}(C^{I'}(t)+
(s-u)D^{I'}(t))+\nonumber \\
	   &   &{1\over\pi}\int_4^\infty {dx\over x^2}\left(
		{s^2\over x-s} {\bf I}^{II'} +
		{u^2\over x-u}C_{su}^{II'}\right)\ A^{I'}(x,t),
\end{eqnarray}
where $A^{I}(x,t)$ 
is the isospin $I$
s-channel absorptive part, $C_{st}$ and $C_{su}$ are the crossing matrices:
\begin{eqnarray*}
	C_{st} = \pmatrix{
	 1/3 & 1 & 5/3 \cr
	 1/3 & 1/2 & -5/6 \cr
	 1/3 & -1/2 & 1/6 \cr
	 },\quad
	C_{su} = \pmatrix{
	 1/3 & -1 & 5/3 \cr
	 -1/3 & 1/2 & 5/6 \cr
	 1/3 & 1/2 & 1/6 \cr
	 }\quad
%
\end{eqnarray*}
and ${\bf I}$ is the identity matrix.
Bose symmetry implies: $C^1(t)=D^0(t)=D^2(t)=0$;
the unknown t-dependent functions
$C^I(t)$ and $D^I(t)$ may be eliminated,
using crossing symmetry, in favor of the S-wave
scattering lengths.

The Roy equations are obtained upon
projecting the resulting dispersion relation onto
partial waves and inserting a partial wave expansion for the absorptive part.
They have been rigorously proved to be valid in the domain
$-4\leq s\leq 60$.
These are a system of coupled integral equations for
partial wave amplitudes of definite isospin $I$ which are related
through crossing symmetry to the absorptive parts of all the
partial waves.  
The Roy equations for the S- and P- waves are
\cite{roy,BGN,BFP}:
\begin{eqnarray}\label{roy_eq}
f^0_0(s) & = & a^0_0 + ( 2 a^0_0 - 5 a^2_0) \frac{s-4}{12}
      		+ \sum_{I'=0}^{2} \sum_{l'=0}^{\infty}\int_{4}^{\infty}dx
		K^{l'I'}_{00}(s,x) \mbox{Im} f^{I'}_{l'}(x),\nonumber\\
f^1_1(s) & = & ( 2 a^0_0 - 5 a^2_0) \frac{s-4}{72}
      		+ \sum_{I'=0}^{2} \sum_{l'=0}^{\infty}\int_{4}^{\infty}dx
		K^{l'I'}_{11}(s,x) \mbox{Im} f^{I'}_{l'}(x),\\
f^2_0(s) & = & a^2_0 - ( 2 a^0_0 - 5 a^2_0) \frac{s-4}{24}
      		+ \sum_{I'=0}^{2} \sum_{l'=0}^{\infty}\int_{4}^{\infty}dx
		K^{l'I'}_{20}(s,x) \mbox{Im} f^{I'}_{l'}(x) \nonumber
\end{eqnarray}
and for all the higher partial waves written as:
\begin{eqnarray*}
f^I_l(s) & = &  \sum_{I'=0}^{2} \sum_{l'=0}^{\infty}\int_{4}^{\infty}dx
		K^{l'I'}_{Il}(s,x) \mbox{Im} f^{I'}_{l'}(x),\quad l\geq 2,
\end{eqnarray*}
where $K^{l'I'}_{lI}(s,s')$ are 
the kernels of the integral equations and have been documented
elsewhere~\cite{BGN}. 
Upon cutting off the integral at a large scale
$\Lambda$ 
and absorbing the contribution of the high energy tail as well as
that of all the higher waves over the entire energy range
into the  
driving terms $d^I_l(s,\Lambda)$ we have:
\newpage 
\begin{eqnarray}\label{roy_eq_drv}
f^0_0(s) & = & a^0_0 + ( 2 a^0_0 - 5 a^2_0) \frac{s-4}{12}
      		+ \sum_{I'=0}^{2} \sum_{l'=0}^{1}\int_{4}^{\Lambda}dx
		K^{l'I'}_{00}(s,x) \mbox{Im} f^{I'}_{l'}(x) \nonumber \\
& & +		d^0_0(s,\Lambda),\nonumber\\
f^1_1(s) & = & ( 2 a^0_0 - 5 a^2_0) \frac{s-4}{72}
      		+ \sum_{I'=0}^{2} \sum_{l'=0}^{1}\int_{4}^{\Lambda}dx
		K^{l'I'}_{11}(s,x) \mbox{Im} f^{I'}_{l'}(x) \nonumber \\
& & +		d^1_1(s,\Lambda),\\
f^2_0(s) & = & a^2_0 - ( 2 a^0_0 - 5 a^2_0) \frac{s-4}{24}
      		+ \sum_{I'=0}^{2} \sum_{l'=0}^{1}\int_{4}^{\Lambda}dx
		K^{l'I'}_{20}(s,x) \mbox{Im} f^{I'}_{l'}(x) \nonumber \\
& & +		d^2_0(s,\Lambda) \nonumber
\end{eqnarray}
and for all the higher partial waves written as:
\begin{eqnarray*}
f^I_l(s) & = &  \sum_{I'=0}^{2} \sum_{l'=0}^{1}\int_{4}^{\Lambda}dx
		K^{l'I'}_{Il}(s,x) \mbox{Im} f^{I'}_{l'}(x)
		+d^I_l(s,\Lambda),\quad l\geq 2.
\end{eqnarray*}
The driving terms themselves for the two lowest partial waves,
the $I=0,2$ S- waves and the $I=1$ P- wave are available in
the literature when $\Lambda=110$ \cite{BFP}. 
[We note that the driving terms for the
$l\geq 2$ partial waves are not documented in the
literature.]  The primary aim of this work is to derive
as completely as possible the information on the S- and P- waves
and that of the numerical fit to the experimental data (described in
a subsequent section) to provide a parametric representation
for the physical S- and P- wave partial waves.

Out of the absorptive parts of the physical S- and P- waves,
one may construct a manifestly crossing
symmetric amplitudes~\cite{BGN,mahoux}:

\begin{eqnarray}\label{eq:new_roy}
\lefteqn{{1\over 32 \pi}T^I(s,t,u)  =} \nonumber \\
&  &\sum_{I'=0}^2
\frac{1}{4}(s {\bf I}^{II'}+ t C_{st}^{II'} + u C_{su}^{II'}) a_0^{I'}+
	\frac{1}{\pi}\int_4^\infty \frac{dx}{x(x-4)} \cdot \nonumber \\
&  &	\left\{\left[
        \frac{s(s-4)}{x-s} {\bf I}^{II'}+ \frac{t(t-4)}{x-t} C_{st}^{II'} +
	\frac{u(u-4)}{x-u} C_{su}^{II'}\right] {\rm Im}
f_0^{I'}(x)  \right.\\ 
			&   & \left.
+3\left[\frac{s(t-u)}{x-s} {\bf I}^{II'} + 
\frac{t(s-u)}{x-t} C_{st}^{II'} +
	\frac{u(t-s)}{x-u} C_{su}^{II'}
\right] {\rm Im}  f_1^{I'}(x)\right\}.  \nonumber
\end{eqnarray}
Our objective may be
met by first rewriting this dispersion relation as:
\begin{small}
\begin{eqnarray}
\lefteqn{{1\over 32 \pi} T^{I}(s,t,u)  = } \nonumber\\
	&   & \sum_{I'=0}^{2} {1\over 4}(s{\bf I}^{II'}+ t C_{st}^{II'}
	      + u C_{su}^{II'})\alpha_0^{I'} + (s^2{\bf I}^{II'} 
	      + t^2 C_{st}^{II'} + u^2 C_{su}^{II'})\gamma_0^{I'}\nonumber \\
	&   & +(s(t-u){\bf I}^{II'} + t(s-u)C_{st}^{II'}+u(t-s)C_{su}^{II'})
	      \beta_1^{I'}  \nonumber \\ 
	&   & +{1\over \pi}\int_4^\infty {dx\over x^3}
	      \left( {s^3\over x-s}{\bf I}{II'}+{t^3\over x-t}C_{st}^{II'}+
	      {u^3\over x-u}C_{su}^{II'}\right) {\rm Im} f_0^{I'}(x) 
	      \nonumber \\
	&   & +{3\over \pi}\int_4^\infty {dx\over x^2 (x-4)}
	      \left( {s^2 (t-u)\over x-s}{\bf I}^{II'}+{t^2 (s-u) \over x-t}
	      C_{st}^{II'}+{u^2 (t-s)\over x-u} C_{su}^{II'}\right){\rm Im}
	      f_1^{I'}(x),\nonumber
\end{eqnarray}
\end{small}
where
$\alpha^I_0$, $\beta^1_1$ and $\gamma^I_0$ are defined in eq.(\ref{abg}).
We now invert the isospin amplitudes eq.~(\ref{eq:amp:iso:def}) to obtain
$A(s,t,u)$ that is constructed out of the 
S- and P- wave absorptive parts:
\begin{eqnarray}
A(s,t,u) & \equiv & {1\over 3} \left(
T^0(s,t,u)-T^2(s,t,u)\right) 
\end{eqnarray}
and is given in eq. (\ref{Adisprecon}).
\newpage

\newpage

\noindent{\bf Table Captions}

\bigskip

\noindent {Table 1.}  The computed values corresponding to the
input $a^0_0$ of $a^2_0, $ $ b^0_0,$ $ b^2_0, $ $  a^1_1, $ $ b^1_1$ and
the computed values of \lb{1}, \lb{2} and \lb{4}.

\bigskip

\noindent {Table 2.} (a) Values of combinations of threshold parameters
corresponding to the fit
of the first line of Table 1 and their one loop values  with the new
$\lb{1}$ and $\lb{2}$; (b) As in (a) for the second line; (c)
As in (a) for the third line

\bigskip

\bigskip

\noindent{\bf Figure Caption}

\bigskip
\noindent{Fig.~1.} 
Results of our fit to the combination
$\delta^0_0-\delta^1_1$ (in degrees) as a function
of energy (MeV): full line $a^0_0 = 0.26$
		dash--dotted line: $a^0_0 = 0.21$, dashed line: $a^0_0 = 0.2$
		dash--double--dotted line: $a^0_0 = 0.19$.
Also shown is the Rosselet data.
\newpage

$$
\begin{array}{||c|c|c|c|c|c|c|c|c|c||}\hline
 & a^0_0 & a^2_0 & b^0_0 & b^2_0 & a^1_1 & b^1_1 & \lb{1} & \lb{2} & \lb{4} \\
\hline
1 & 0.19 & -0.040 & 0.238 & -0.074 & 0.035 & 0.006 & -1.80 & 4.98 & 
0.87   \\
2 & 0.20 & -0.037 & 0.237 & -0.074 & 0.035 & 0.006 & -1.69 & 4.97 &
1.09   \\
3 & 0.21 & -0.035 & 0.238 & -0.075 & 0.035 & 0.006 & -1.58 & 4.96 &
1.30   \\ \hline
\end{array}
$$
\begin{center}
Table 1
\end{center}

\bigskip

$$
\begin{array}{||c|c|c||}\hline 
 & \mbox{Roy equations} & \mbox{One loop formula} \\ \hline
18 a^1_1 - (2 a^0_0-5 a^2_0) & 0.0459 & 0.0472 \\
18 b^1_1  & 0.1079 & 0.0788 \\
a^0_2 & 15\times10^{-4} &13\times10^{-4}  \\
a^2_2 & 0.6\cdot 10^{-4}& 0.4\cdot 10^{-4}\\
3 b^0_0  - (2 a^0_0-5 a^2_0) & 0.1326& 0.0924\\
6 b^2_0 + (2 a^0_0-5 a^2_0) & 0.1378 & 0.1370\\ 
a_2 & 0.0005 & 0.0005 \\
b_0 & 0.030 & 0.026 \\
2a^0_0-5a^2_0 & 0.580  & 0.561  \\ \hline
\end{array}
$$

\begin{center}
Table 2 (a)
\end{center}

$$
\begin{array}{||c|c|c||}\hline 
 & \mbox{Roy equations} & \mbox{One loop formula} \\ \hline
18 a^1_1 - (2 a^0_0-5 a^2_0) & 0.0410 & 0.0436\\
18 b^1_1  & 0.1066 & 0.0753 \\
a^0_2 & 15\times10^{-4} &13\times10^{-4} \\
a^2_2 & 0.7\cdot 10^{-4}& 1.0\cdot 10^{-4}\\
3 b^0_0  - (2 a^0_0-5 a^2_0) & 0.1269 & 0.0989\\
6 b^2_0 + (2 a^0_0-5 a^2_0) & 0.1423 & 0.1400\\
a_2 &  0.0005 & 0.0005 \\
b_0 & 0.030 & 0.027 \\ 
2a^0_0-5a^2_0 & 0.585  & 0.563 \\ \hline
\end{array}
$$

\begin{center}
Table 2 (b)
\end{center}

$$
\begin{array}{||c|c|c||}\hline 
 & \mbox{Roy equations} & \mbox{One loop formula} \\ \hline
18 a^1_1 - (2 a^0_0-5 a^2_0) & 0.0360 & 0.0424\\
18 b^1_1  & 0.1053& 0.0740\\
a^0_2 & 15\times10^{-4} & 13\times10^{-4}\\
a^2_2 & 0.7\cdot 10^{-4}& 1.0\cdot 10^{-4}\\
3 b^0_0  - (2 a^0_0-5 a^2_0) & 0.1200 & 0.1008\\
6 b^2_0 + (2 a^0_0-5 a^2_0) & 0.1472 & 0.1407\\ 
a_2 & 0.0005 & 0.0005 \\
b_0 & 0.030 & 0.027 \\ 
2a^0_0-5a^2_0 & 0.595 & 0.567  \\ \hline
\end{array}
$$

\begin{center}
Table 2 (c)
\end{center}

\newpage

\begin{figure}[h]
\epsfxsize=8cm
\centerline{\epsffile{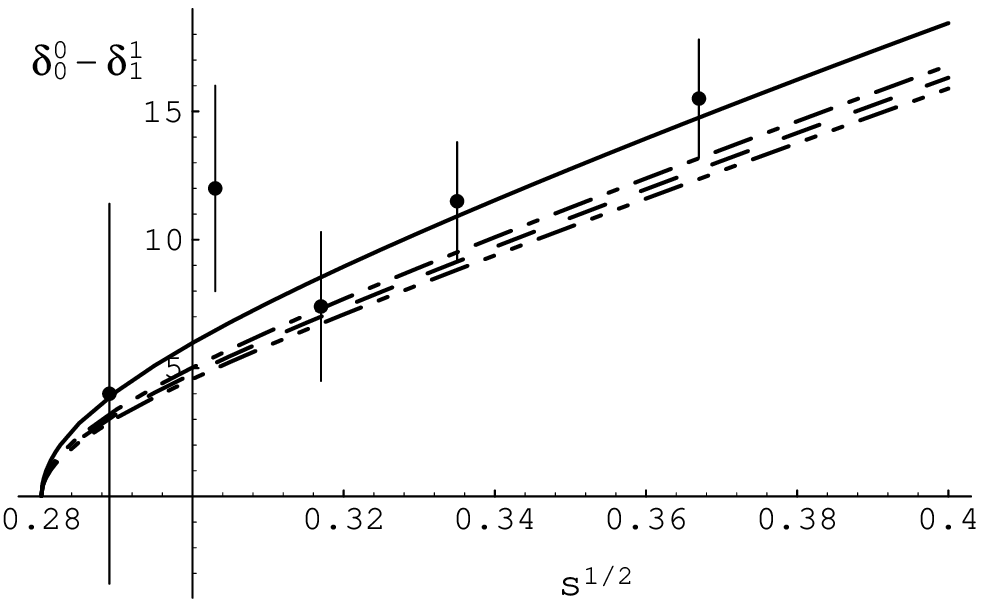}}
\end{figure}
\centerline{Fig.~1}

\end{document}